\documentclass[pre,aps,twocolumn, superscriptaddress]{revtex4}

\usepackage{color}
\usepackage{epsfig}
\usepackage{amssymb}
\usepackage{amsmath}

\begin{document}

\title{Application of underdamped Langevin dynamics simulations for
  the study of diffusion from a drug-eluting stent}

\author{Shaked Regev} 
\affiliation{Department of Biomedical
Engineering, Ben-Gurion University of the Negev, Be'er Sheva 85105,
Israel} 

\author{Oded Farago} 
\affiliation{Department of Biomedical
Engineering, Ben-Gurion University of the Negev, Be'er Sheva 85105,
Israel} 
\affiliation{Ilse Katz Institute for Nanoscale Science and
Technology, Ben-Gurion University of the Negev, Be'er Sheva 85105,
Israel}

\begin{abstract}

  We use a one-dimensional two layer model with a semi-permeable
membrane to study the diffusion of a therapeutic drug delivered from a
drug-eluting stent (DES). The rate of drug transfer from the stent
coating to the arterial wall is calculated by using underdamped
Langevin dynamics simulations. Our results reveal that the membrane
has virtually no delay effect on the rate of delivery from the
DES. The work demonstrates the great potential of underdamped Langevin
dynamics simulations as an easy to implement, efficient, method for
solving complicated diffusion problems in systems with a
spatially-dependent diffusion coefficient.

\end{abstract}

\maketitle

\section{Introduction}
\label{sec:intro}

Arterial stents are indispensible in the treatment of coronary artery
disease (CAD) and more specifically stenosis, the abnormal narrowing
of blood vessels \cite{renal}. These stents are frequently implanted
in arteries where blood flow has become precariously impeded. In
recent decades, they have revolutionized the treatment of stenosis by
providing a safer alternative to coronary surgery. In addition to
diminishing the risk of major surgery complications, stents also
facilitate recovery and avoid administering general anesthesia to
patients \cite{first}. However, arterial stents have only been able to
reduce the instances of recurring stenosis, or restenosis, to 20-30\%,
compared to 30-40\% in coronary surgery \cite{clinic}.

In an effort to further curtail these rates, drug-eluting stents (DES)
were introduced. A DES is a stent that uses programmed pharmacokinetics
to release anti-proliferative pharmaceuticals into the arterial
wall. It is comprised of a metallic strut coated with a polymeric
matrix or gel that encapsulates the therapeutic drug~\cite{patho}. The
drug reduces smooth muscle cell growth and prevents an inflammatory
response - two predominant causes of in-stent-restenosis and
neo-intima proliferation~\cite{patho}. Stents coated in
anti-proliferative agents have mitigated instances of restenosis to
roughly 5\% in clinical trials and are FDA approved \cite{hist}.

Understanding the rate at which drugs are transported through the
arterial tissue is crucial for stent design, and as such has been
studied extensively
\cite{review,stent,pharm,curve,marker,design,runge,Laplace,noFick,genmod,anisotropy,siam,nlayers,siam2,compmech,hydro,esaim,model2,stent2,exper,exact,approx,novel,mass,multi,circ1,circ2,binding}.
It has been identified that the major mechanism of drug transfer from
the coating is diffusion through the arterial
wall~\cite{lovich}. Thus, advective forces arising, e.g., from blood
circulation in the arteries have been ignored in many models.  The
simplest model describing the system is based on the solution of a
diffusion equation in a two-layer one-dimensional system.
One-dimensional models represent a mathematical idealization of the
three-dimensional stent geometry; nevertheless, they dominate the
theoretical literature on the subject because the drug release is
predominately along the normal direction to the stent axis whose
dimension (the stent thickness) is much smaller then the lateral
dimension (the stent radius). More complex one-dimensional models give
weight to other phenomena and factors such as chemical reactions
between the drug and the arterial wall
\cite{noFick,genmod,anisotropy,design,review,runge,stent2,multi,binding},
directed advection of the drug
\cite{noFick,genmod,anisotropy,marker,review,nlayers,compmech,esaim,model2,stent2,mass,multi,binding},
cell metabolism \cite{compmech,mass}, and the drug topcoat membrane
permeability \cite{stent,genmod,siam2,hydro,binding}. These phenomena
and factors amend the partial differential equations (PDEs) that
describe the transport of the drug. They also modify the boundary
conditions between the layers and often require the introduction of
additional layers
\cite{multi,stent2,Laplace,model2,nlayers,curve,exact,approx,hydro}.
Some studies consider more complex two- \cite{curve,noFick,novel} and
even three-dimensional \cite{anisotropy,pharm,siam,compmech,hydro}
geometries. The PDEs are often solved by separation of variables, or
numerically through some kind of a discretization scheme, e.g., finite
elements, finite differences, and the marker cell method. Noteworthy
exceptions include analytical Laplace transform solutions
\cite{Laplace}, Boltzmann reductions~\cite{design}, and numerically
solved Voltera integral equations~\cite{siam2}. Experimental data is
also available \cite{circ1,circ2,exper}, and has been used to test and
validate theoretical models.

Layered systems, where in each region the diffusion coefficient takes
a different value, constitute a subclass of a more general class of
systems with spatially varying diffusivity: $D=D(x)$. Diffusion in
such systems is described by Fick's second law
\begin{eqnarray}
\frac {\partial P(x,t)}{\partial t} =-\frac {\partial
  J\left(x,t\right)}{\partial x}=\frac{\partial}{\partial
  x}\left(D\left(x\right) \frac{\partial P(x,t)}{\partial x}\right),
\label{eq:diffusion}
\end{eqnarray}
where $P(x,t)$ is the probability of being at coordinate $x$ at time
$t$, and $J(x,t)=-D(x)\partial_x P(x,t)$ is the probability flux. An
alternative route for calculating $P(x,t)$ is to numerically
integrate the corresponding Langevin equation \cite{langevin}
\begin{eqnarray}
m\frac {dv}{dt}=-\alpha(x) v+\beta\left(x\left(t\right)\right),
\label{eq:langevin}
\end{eqnarray}
where $m$ and $v$ denote, respectively, the mass and velocity of a
diffusing particle. Langevin's equation describes Newtonian dynamics
under the action of two effective forces: friction [first term on the
  r.h.s.~of Eq.~(\ref{eq:langevin})] and stochastic thermal noise
(second term). The friction coefficient $\alpha(x)$ in
(\ref{eq:langevin}) is related to the diffusivity $D(x)$ in
(\ref{eq:diffusion}) via Einstein's relation $\alpha(x)=k_BT/D(x)$
\cite{einstein,gjf4}, where $T$ is the temperature of the system and
$k_B$ is Boltzmann's constant. The stochastic noise, $\beta$, is
chosen from a Gaussian distribution with zero mean $\langle
\beta(x(t)) \rangle=0$, and delta-function auto-correlation $\langle
\beta(t) \beta(t') \rangle =2k_BT\alpha(x(t))
\delta(t-t')$\cite{gjf4,gjf3}. The probability density function (PDF)
of the particle, $P(x,t)$, can be computed from an ensemble of
simulated trajectories, where the initial position of the particle at
each trajectory is chosen from the intial PDF $P(x,0)$.

The accuracy of the Langevin dynamics simulation method for computing
$P(x,t)$ depends largely on the quality of the discrete-time numerical
integrator. Here (as in previous works \cite{proj}), we use the
efficient and robust method of Gr\o nbech-Jensen and Farago (G-JF)
\cite{gjf1,gjf2} in combination with the ``inertial'' convention
\cite{gjf3,gjf4} for choosing the value of $\alpha$. In the absence of
conservative forces from a potential energy gradient (which is the
case discussed throughout this paper), the G-JF integrator employs the
following set of equations
\begin{eqnarray}
  x^{n+1}&=&x^n+bdtv^n+\frac{bdt}{2m}\beta^{n+1},
  \label{eq:gjfx0}\\
  v^{n+1}&=&av^n+\frac{b}{m}\beta^{n+1},
  \label{eq:gjfv0}
\end{eqnarray}
to advance the coordinate $x^n=x(t_n)$ and velocity $v^n=v(t_n)$ by
one time step from time $t_n=ndt$ to $t_{n+1}=t_n+dt$. In the above
G-JF equations (\ref{eq:gjfx0})-(\ref{eq:gjfv0}), $\beta^{n+1}$ is a
Gaussian random number satisfying
\begin{eqnarray}
\left\langle \beta^n\right\rangle=0\ ;\  \left\langle
\beta^n\beta^l\right\rangle=2\alpha k_BT dt\delta_{n,l},
\label{eq:noise0}
\end{eqnarray}
and the dampening coefficients of the algorithm are
\begin{eqnarray}
  b=\left[1+\left(\alpha dt/2m\right)\right]^{-1}\ ;
  \ a=\left[1-\left(\alpha dt/2m\right)\right]b.
  \label{eq:constants0}
\end{eqnarray}

Since the friction varies in space, the integrator must be
complemented with a convention for choosing the value of $\alpha$ to
be used in Eqs.(\ref{eq:noise0}) and (\ref{eq:constants0}) at each
time step. The ambiguity concerning the appropriate choice of $\alpha$
is known in the literature as the ``It\^{o}-Stratonovich
dilemma''\cite{ito,strat,dilemma}. Here, we use the recently proposed
inertial convention \cite{gjf3,gjf4} that assigns to $\alpha$ the
value of the spatial average of $\alpha(x)$ along the inertial
trajectory from $x^n$ to $\tilde{x}^{n+1}=x^n+v^ndt$
\begin{equation}
\bar{\alpha}\equiv\frac{\int_{x^n}^{\tilde{x}^{n+1}} \alpha (x)
  dx}{\tilde{x}^{n+1}-x^n}=\frac{A(\tilde{x}^{n+1})-A(x^n)}{\tilde{x}^{n+1}-x^n},
\label{eq:avgalpha0}
\end{equation}
where $A(x)$ is the primitive function of $\alpha(x)$. We have
previously demonstrated that while any reasonable convention for
choosing $\alpha$ yields the correct PDF at the limit of
infinitesimally small time step $dt\rightarrow 0$, the inertial
convention gives very accurate results even for relatively large time
steps. This property of the inertial convention stems from the fact
that at small time scales, $dt\ll m/\alpha$,  the inertial trajectory
serves as the leading approximation to the actual path of the
particle.

\section{Two-layer systems}
\label{sec:twolayer}

\begin{figure}[t]
\centering\includegraphics[width=0.525\textwidth]{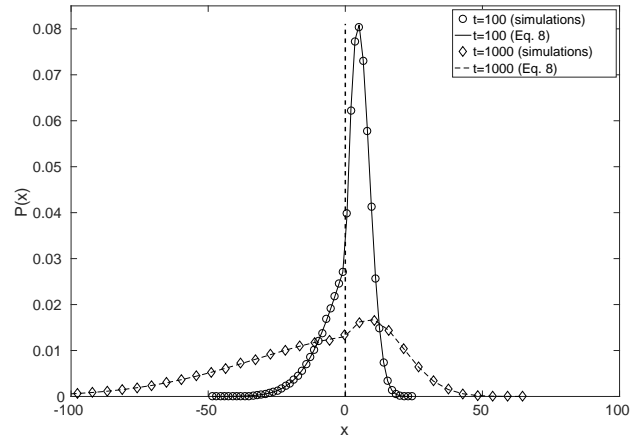}
\caption{The PDF of the two-layer problem (the vertical line at $x=0$
  represents the interface).  Open circles and the solid line denote,
  respectively, the results of the Langevin dynamics simulations and
  the analytical solution Eq.~(\ref{eq:2layer}) at $t=100$. Similarly,
  the open diamonds and the dashed line denote results for $t=1000$.}
\label{fig:fig1b}
\end{figure}

Assuming in Eq.~(\ref{eq:diffusion}) that $0<D(x)<\infty$
($-\infty<x<\infty$), the resulting $P(x,t)$ (for $t>0$) must be a
continuous function for any initial condition. The flux
$j(x,t)=-D(x)\partial_x P(x,t)$ must also be continuous if no source
or sink of probability are present in the system. These properties of
the PDF also apply to layered systems where $D$ is piece-wise
constant. In this paper, we consider a simple one-dimensional
two-layer model for a DES. Before arriving at that model, we first
consider the simplest two-layer system, where $D(x)=D_1$ for $x<0$ and
$D(x)=D_2$ for $x>0$. As just stated above, both the PDF and the flux
must be continuous, including at $x=0$, which sets the boundary
conditions at the interface between the layers. Assuming that a
particle is initially located at $x=x_0>0$, i.e.,
$P(x,0)=\delta(x-x_0)$, then the {\em normalized} (i.e., satisfying
$\int_{-\infty}^{\infty} P(x,t)dx=1$) solution of (\ref{eq:diffusion})
for $t>0$ is given by
\begin{equation}
  P(x,t)=\left\{\begin{array}{ll}
 \frac{A}{\sqrt{4\pi D_1t}}e^{-\frac{(x-x_1)^2}{4D_1t}} & x<0 \\
 \frac{1}{\sqrt{4\pi D_2t}}e^{-\frac{(x-x_0)^2}{4D_2t}}+\frac{B}
    {\sqrt{4\pi D_2t}}e^{-\frac{(x+x_0)^2}{4D_2t}} & x>0\ ,
\end{array} \right.
  \label{eq:2layer}
\end{equation}
with $x_1=x_0\sqrt{D_1/D_2}$, $A=2/(1+\sqrt{D_2/D_1})$, and
$B=(1-\sqrt{D_1/D_2})/(1+\sqrt{D_1/D_2})$. This solution can be
interpreted as follows: For $x>0$, the PDF is the outcome of two
diffusion processes associated with (i) the original particle which
has a unity weight and is located at $x=x_0$, and an image particle of
weight $B$ located at $x=-x_0$. For $x<0$, the PDF represents
diffusion of an image particle of weight $A=1-B$, which is located at
$x=x_1$. Notice that for $D_1=D_2$, we have $x_1=x_0$, $A=1, B=0$,
which reduces Eq.~(\ref{eq:2layer}) to the well-known Gaussian
solution for a particle diffusing in an infinite space with constant
diffusivity.  Fig.~\ref{fig:fig1b} compares the PDF (\ref{eq:2layer})
for $D_1=1$, $D_2=0.1$, and $x_0=4$ (lines) to the results of Langevin
dynamics simulations based on the above described G-JF integrator with
the inertial convention (symbols). In the simulations, we set $m=1$
and $k_BT=1$. Thus, the friction coefficients in the layers are given
by $\alpha_1=k_BT/D_1=1$ and $\alpha_2=k_BT/D_2=10$. We set the time
step to $dt=0.1$. For both layers, this value of $dt$ satisfies the
condition $dt\leq 2m/\alpha$, which has been assumed in the
application of the inertial convention for choosing $\bar{\alpha}$
(see discussion at the end of section
\ref{sec:intro}). Fig.~\ref{fig:fig1b} depicts the PDF at $t=100$ and
$t=1000$, based on $2.5\times 10^5$ trajectories. The total CPU time
of the simulations was 3 minutes on a PC. The agreement with
Eq.~(\ref{eq:2layer}) is excellent. Interestingly, we obtained almost
identical results with $dt=4$, which does not satisfy the condition
for inertial dynamics. In the latter case, the total duration of the
simulations was 5 seconds.

Two interesting limiting cases may be considered: For
$D_2/D_1\rightarrow \infty$ ($D_2=D$, $D_1\rightarrow 0$), we have $B
\rightarrow 1$, and
\begin{equation}
  P(x,t)=\begin{array}{ll}
  \frac{1}{\sqrt{4\pi
      D_2t}}e^{-\frac{(x-x_0)^2}{4D_2t}}+\frac{1}{\sqrt{4\pi
      D_2t}}e^{-\frac{(x+x_0)^2}{4D_2t}} & (x>0),
  \end{array} 
\label{eq:reflecting}
\end{equation}
which is the solution of the same problem with a reflecting wall at
the origin. To simulate Langevin dynamics in the presence of a
reflecting wall (located, without loss of generality, at $x=0$), we
follow the trajectory of the particle as computed by the Langevin
integrator. If it crosses the wall, i.e., when $x^{n+1}<0$, we simply
relocate the particle at $-x^{n+1}>0$, and reverse the velocity from
$v^{n+1}$ to $-v^{n+1}$. The other limiting case is $D_2/
D_1\rightarrow 0$ ($D_2=D$, $D_1\rightarrow \infty$). Here, we have $B
\rightarrow -1$, and
\begin{equation}
  P(x,t)=\begin{array}{ll} \frac{1}{\sqrt{4\pi
      D_2t}}e^{-\frac{(x-x_0)^2}{4D_2t}}-\frac{1}{\sqrt{4\pi
      D_2t}}e^{-\frac{(x+x_0)^2}{4D_2t}} & (x>0), 
  \end{array}
 \label{eq:absorbing}
\end{equation}
which is the solution of the same problem with an absorbing wall at
the origin. When one simulates Langevin dynamics with absorbing
boundaries, the trajectory is terminated when it crosses the
boundary. Obviously, the total probability is not conserved but rather
diminishes with time.

\section{Semi-permeable membrane}
\label{sec:semiper}

Drug-eluting stents include a topcoat that influences
the rate of drug release to the artery. Since the
topcoat layer is thin compared to the dimensions of
the stent and the arterial wall, it can be included in a DES two-layer
model as a semi-permeable membrane boundary that controls the
transition rate from the coating (first layer) to the
artery (second layer). Let us first discuss the problem of diffusion
across a semi-permeable membrane within the general context of
two-layer systems.  We denote by $P_{\rm cross}$ the probability that,
within a small time interval $dt$, a drug molecule arriving to the
boundary from the first layer crosses it to the second layer. In the
opposite direction (from the second layer back to the first), the
crossing probability is unity. A semi-permeable membrane can be
incorporated in Langevin dynamics simulations of layered systems in
the following manner: Let us first consider the simpler case where the
friction coefficient $\alpha$ is the same on both sides of the
membrane, which is located at $x=x_m$. We follow the particle until it
crosses the membrane ($x^n<x_m$ and $x^{n+1}>x_m$). In order to decide
whether the particle should cross the boundary or be reflected from
it, we draw a random number, $R$, from a uniform distribution between
zero and one. We accept the new coordinate and velocity
$\left(x^{n+1},v^{n+1}\right)$ if $R<P_{\rm cross}$ (crossing), and
reverse them to $\left(2x_m-x^{n+1},-v^{n+1}\right)$ if $R>P_{\rm
  cross}$ (reflection). If $\alpha$ varies across the membrane, the
algorithm is only slightly more complicated: The new coordinate and
velocity $(x^{n+1},v^{n+1})$ are computed assuming that the friction
varies in space, i.e., with the inertial convention for the average
friction during the time step [Eq.~(\ref{eq:avgalpha0})]. The step is
accepted for $R<P_{\rm cross}$ and is changed to a reflection step for
$R>P_{\rm cross}$. In the latter case, one must take into account the
fact that the particle has not passed the membrane and, therefore, the
friction coefficient along its trajectory remains constant. Therefore,
the new coordinate and velocity are recalculated using the G-JF
algorithm, with the already chosen value of $\beta^{n+1}$, but with
the friction coefficient at the initial coordinate $x^n<x_m$. The
recalculated $(x^{n+1},v^{n+1})$ are accepted if the particle has not
crossed the membrane ($x^{n+1}<x_m$), and changed to
$\left(2x_m-x^{n+1},-v^{n+1}\right)$ if it has.

\begin{figure}[t]
\centering\includegraphics[width=0.5\textwidth]{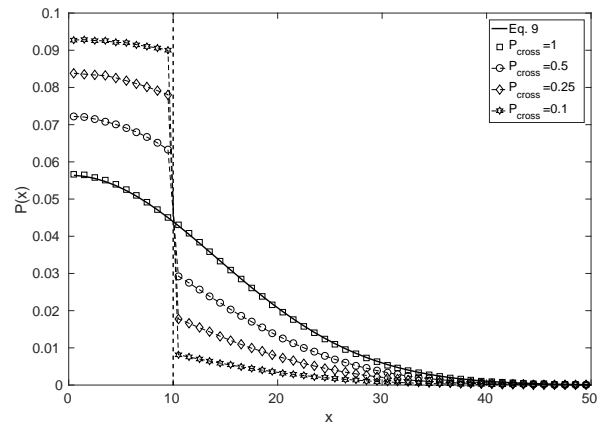}
\caption{The PDF at $t=100$ of a particle starting next to a
  reflecting wall at the origin, and diffusing through a
  semi-permeable membrane at $x=10$ (represented by the vertical
  line), with crossing probability $P_{\rm cross}$.  Squares, circles,
  diamonds and stars denote the results of the Langevin dynamics
  simulations for $P_{\rm cross}=1,0.5,0.25,0.1$, respectively. The
  solid line shows the solution for the case with no membrane [see
    Eq.~(\ref{eq:reflecting})].}
\label{fig:fig2b}
\end{figure}

Fig.~\ref{fig:fig2b} shows simulation results for the PDF of a
particle, initially located right next at a reflecting wall at $x=0$,
and diffusing in a medium with constant $D=1$ that has a
semi-permeable membrane at $x_m=10$. The PDF at $t=100$ is plotted for
various values of $P_{\rm cross}$. The PDF is discontinuous at $x_m$
because the membrane interferes with the diffusion rates. However, the
membrane is not a probability sink or source and, therefore, the
probability flux, $j(x)=-D\partial_x P(x,t)$, arriving at the membrane
from left ($x\rightarrow x_m-$) is the same as the flux leaving the
membrane to the right ($x\rightarrow x_m+$). This last feature of the
PDF is visually apparent in Fig.~\ref{fig:fig2b}. Since the membrane
(partially) blocks the drug flow only from left to right, the
probability density drops across the membrane, i.e.,
$P(x_m-,t)>P(x_m+,t)$. The data in Fig.~\ref{fig:fig2b} suggests that
the following relation holds: $P_{\rm
  cross}=P(x_m+,t)/P(x_m-,t)$. This relation is anticipated from the
continuity of the flux and the fact that for each $P_{\rm cross}$
molecules crossing the membrane from left to right, one molecule
passes from right to left. For $P_{\rm cross}=1$, the membrane is
effectively non-existent because it does not impede the diffusion. In
this limit, $P(x,t)$ becomes continuous at $x_m$ and is given by
Eq.~(\ref{eq:reflecting}) with $x_0=0$ (solid line in
Fig~\ref{fig:fig2b}).

\section{Two-layer stent model}

\subsection{The model}

\begin{figure}[t]
  \centering\includegraphics[width=0.5\textwidth]{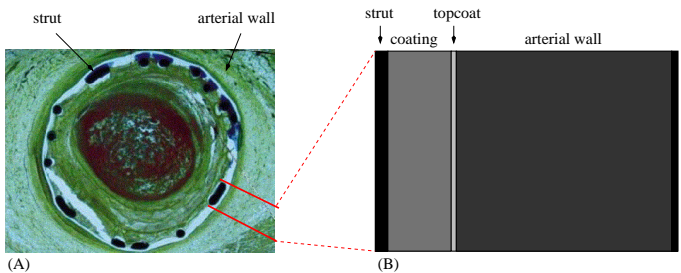}
  \caption{(A) A cross section of a stented artery, and (B)
    the corresponding one-dimensional two-layer model for drug release
    from the stent to the artery.}
\label{fig:fig3b}
\end{figure}

\begin{table}[t]
\begin{center}
  \begin{tabular}{ | c | c | c | c | }
    \hline
parameter & physical & normalized & simulation  \\
\ & value & value & value \\ \hline     
$L_1$ & $5\times10^{-6} {\rm\, m}$ & 1 & 1 \\ \hline
$L_2$ & $10^{-4} {\rm\, m}$ & 20 & 20 \\ \hline
$D_1$ & $10^{-17} {\rm\, m^2/s}$ & 1/700 & 1/700  \\ \hline
$D_2$ & $7\times10^{-15} {\rm\, m^2/s}$ & 1 & 1 \\ \hline
$k_BT$ & $4.3\times10^{-21} {\rm\, J}$ & 1 & 1 \\ \hline
$m$ &  $1.5\times10^{-24} {\rm\, kg}$ & $6.8\times10^{-19}$ & 1 \\ \hline
$\tau_1$ & $2.5\times10^6{\rm\, s}$ & 700 & 700 \\ \hline
$\tau_2$ & $1.4\times10^6 {\rm\, s}$ & 400 & 400 \\ \hline
$\tau_m$ & $3.5\times10^{-18} {\rm\, s}$ & $9.7\times10^{-22}$ & 1/700 \\ \hline
  \end{tabular}
  \caption{Model parameters and their physical (second column),
    normalized (third column), and simulation (fourth column) values.}
  \label{tbl:param}
\end{center}
\end{table}

Fig.~\ref{fig:fig3b} (A) shows a cross section of a stented artery
with the struts in contact with the arterial
wall. Fig.~\ref{fig:fig3b}(B) illustrates the mapping of the system to
the simple one-dimensional model discussed herein. The model consists
of two layers representing, respectively, the coating and the arterial
wall. The first layer (coating), which has length $L_1$, is bounded
between the stent strut ($x=-L_1$) and the stent topcoat ($x=0$). The
former is a reflecting boundary, while the latter is a thin
semi-permeable membrane. The second layer (the arterial wall) is of
length $L_2$. It is bounded between the topcoat and the adventitial
side of the arterial wall ($x=L_2$), which is modeled as an absorbing
boundary since the drug arriving at the end of the artery is lost in
the tissues adjacent to the adventitia. The first and second layers
have diffusion coefficients $D_1$ and $D_2$, respectively. We denote
by $m$ the mass of a drug molecule, and use the thermal energy $k_BT$
as the energy scale of the problem.

Typical values for the system parameters are given in the second
column of Table~\ref{tbl:param} in MKS units.  In the third column,
the same parameters are given in normalized units, where $L_1=1$,
$D_2=1$, and $k_BT=1$. Table~\ref{tbl:param} also gives the physical
and normalized values of the time scales $\tau_i=L_i^2/D_i$ ($i=1,2$),
which are the characteristic diffusion times in each layer. Notice the
interesting feature that these two time scales are of the same order
of magnitude.

\subsection{Using a fictitious mass}
\label{sec:fictmass}

 Another characteristic time scale appearing in table~\ref{tbl:param}
 is $\tau_m=mD/k_BT$, which is the crossover time from inertial to
 diffusive Langevin dynamics. This time scale, to be henceforth
 referred to as the {\em ballistic time}\/, is very important from a
 computational perspective since Langevin dynamics simulations with
 the inertial interpretation must be run with time step $dt<\tau_m$ in
 order to appropriately simulate the transition statistics between the
 layers. In multi-layers systems, the limit on $dt$ is set by the most
 viscous layer with the smallest diffusion coefficient and the
 smallest ballistic time. Therefore, in table~\ref{tbl:param}, we only
 give $\tau_m$ of the first layer. For the stent model, we have
 $\tau_m\sim 10^{-18}$ sec, which is more than 24 orders of magnitude
 smaller than the diffusion times $\tau_1$ and $\tau_2$. This poses a
 huge computational challenge, as the aim of the study is to measure
 quantities like the rate of drug transfer to the bloodstream, which
 require simulations on time scales comparable to
 $\tau=\tau_1+\tau_2$. This implies that each simulated trajectory
 requires, at least, $10^{24}$ time-steps, which would make the
 simulations prohibitively time consuming. More specifically,
 simulations of just $10^4$ trajectories would require $~10^{13}$
 years (!) of CPU time on a state of the art PC.

 The key to circumvent this outstanding computational problem is to
 notice that the diffusion equation (\ref{eq:diffusion}) and
 Langevin's equation (\ref{eq:langevin}) yield the same long-time
 probability distributions if Einstein's relation $D(x)\alpha(x)=k_BT$
 is satisfied. The mass of the particle, which only appears in
 Eq.~(\ref{eq:langevin}), is unimportant for this relationship between
 the two equations to hold. Therefore, if we are only interested in
 obtaining the PDF at time scales comparable to $\tau$, we can use a
 fictitious mass which is much larger than the actual mass and, thus,
 artificially increase the ballistic time $\tau_m$. Of course, the
 artificial ballistic time must still be much smaller than the
 diffusion time, but $\tau_m$ does not need to be {\em negligibly}\/
 smaller than $\tau$. In the fourth column of table~\ref{tbl:param} we
 give the normalized values of the model parameters used in our
 Langevin dynamics simulations of the DES model. The simulations
 values are identical to the normalized values for $L_i$, $D_i$, and
 $k_BT$, but the mass in the simulations is set to unity, i.e., about
18 orders of magnitude larger than the actual
 normalized mass of a drug molecule. This narrows the gap between the
 ballistic time and diffusion times to 5-6 orders of magnitude, and
 makes computationally feasible simulations of hundreds of thousands
 of trajectories with integration time-step $dt\ll\tau_m$.

 \subsection{Membrane permeability}
 \label{sec:perm}

 In section \ref{sec:semiper} we defined $P_{\rm cross}$ to be the
 left-to-right crossing probability of the membrane in one time
 step. Since the time step of the simulations is typically several
 orders of magnitude smaller than the physical time scales of interest
 ($dt\ll\tau$), the crossing probability $P_{\rm cross}$ will
 generally be very small, which would considerably slow down the
 simulations. To circumvent this problem, as well as the problem
 arising from the difficulty to accurately estimate the crossing
 probability, we replace $P_{\rm cross}$ with a different measure for
 the membrane permeability - the permeation time $T$. The latter can
 be related to $P_{\rm cross}$ by noticing that each time the particle
 attempts to cross the membrane, it has probability $P_{\rm cross}$ to
 cross it and probability $1-P_{\rm cross}$ to be reflected. Thus, the
 probability of passing the membrane at the $k$-th attempt is $P_{\rm
   cross}(1-P_{\rm cross})^{k-1}$, and the corresponding crossing time
 (measured from the time of the first crossing attempt) is
 $T_k=(k-1)\Delta$, where $\Delta$ is the return time between
 successive crossing attempts. This argument demonstrates that the
 permeation times are geometrically distributed, with an average
 crossing time $\langle T_k\rangle=\Delta (1-P_{\rm cross})/P_{\rm
   cross}$. In reality, $\Delta$ is obviously not fixed but, itself,
 follows a certain continuous distribution. Therefore, the actual
 permeation time does not follow a discrete geometric distribution,
 but its continuous exponential analogue
\begin{eqnarray}
  p(t)=\frac{1}{T}\exp\left(-t/T\right),
\label{eq:exponential}
\end{eqnarray}
where $T$ is the characteristic permeation time. Similarly to its
discrete counterpart $\langle T_k\rangle$, $T=\Delta (1-P_{\rm
  cross})/P_{\rm cross}$, but here $\Delta$ denotes the {\em
  average}\/ time between successive attempts. To account for the
delay effect of a semi-membrane in simulations, we follow the
trajectory of the particle assuming that the membrane is
fully-permeable (i.e., as if there is no membrane present). At each
crossing of the membrane we draw a random waiting time from the
exponential distribution (\ref{eq:exponential}), and this delay time
is simply added to cumulative time of the dynamics.

\section{Results}

The quantity of most interest for therapeutic applications is the rate
of drug transfer from the stent coating to the arterial tissue. This
can be characterized by the fractions $\Pi_1(t)$ and $\Pi_2(t)$ of
drug present in the coating (layer 1) and the arterial wall (layer 2),
respectively. These quantities are related to the PDF via
\begin{eqnarray}
  \Pi_1(t)&=&\int_{-L_1}^{0} P(x,t)dx,\label{eq:pi1}\\
  \Pi_2(t)&=&\int_{0}^{L_2} P(x,t)dx.
\end{eqnarray}
In simulations, each trajectory is terminated upon arrival to the
arterial wall boundary, which implies that $\Pi_1(t)+\Pi_2(t)$ is a
monotonic function decreasing from unity at $t=0$ to zero at
$t\rightarrow \infty$. $\Pi_1(t)$ and $\Pi_2(t)$ are simply identified
with the fraction of trajectories that at time $t$ arrive at some
point within the coating and arterial wall layers, respectively.

\begin{figure}[t]
  \centering\includegraphics[width=0.5\textwidth]{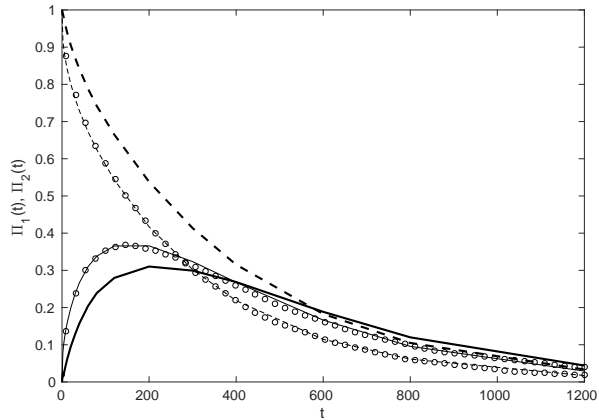}
  \caption{The fraction of drug in the coating (dashed curves),
    $\Pi_1$, and in the arterial wall (solid curves), $\Pi_2$, as a
    function of time. The thin and thick curves show simulation
    results for a DES with semi-permeable membranes having,
    respectively, characteristic permeation times $T=0$ (no membrane)
    and $T=10$. The circles represent the results of ref.~\cite{stent}
    for the same DES model with identical model parameters (see
    footnote \cite{footnote1})}.
\label{fig:fig4b}
\end{figure}

Fig.~\ref{fig:fig4b} shows our computational results for $\Pi_1$
(dashed curves) and $\Pi_2(t)$ (solid curves) as a function of the
time $t$. The results are based on $2.1\times 10^5$ trajectories
simulated with time step $dt=10^{-4}$ which meets the requirement
$dt\ll\tau_m$ where $\tau_m=1/700$ is the ballistic time (see
table~\ref{tbl:param}). The initial coordinates of these trajectories
are chosen from a uniform distribution between $x=0$ and $x=L_1$,
which reflects the initial uniform distribution of the drug within the
coating layer. The thin curves in Fig.~\ref{fig:fig4b} depict results
for the DES model without a semi-permeable membrane, i.e., with
vanishing permeation time $T=0$ between the layers. Not surprisingly,
$\Pi_1$ decreases monotonically from 1 to 0 on a time scale $t\sim
10^3$ (roughly a month and a half in physical units), which is
comparable to the sum of diffusion times in the coating and the
arterial wall $\tau=\tau_1+\tau_2$. $\Pi_2$ increases from an initial
value of 0, arrives to a maximum value for $t\lesssim 200$, and then
monotonically decreases to 0 at larger times. The behavior of $\Pi_2$
reflects net drug transfer from the stent coating to the arterial wall
at the beginning of the process, followed by gradual loss of drug at
longer times, due to absorption in the tissues adjacent to the
adventitia. The thick curves in Fig.~\ref{fig:fig4b} depict results
for the same model but with a semi-permeable membrane characterized by
permeation time $T=10$. The results for $\Pi_1$ and $\Pi_2$ in this
case exhibit trends similar to those observed in the $T=0$
simulations. Importantly, we again observe a decrease in the amount of
drug left in the stent as a function of time and a non-monotonic
behavior of the amount of drug found in the artery.

The open circles in Fig.~\ref{fig:fig4b} depict the results derived in
ref.~\cite{stent} for the same DES model with identical model
parameters (see footnote \cite{footnote1}). These results have been
obtained by numerically solving the diffusion equation at each layer,
subject to a similar initial condition at $t=0$, and boundary
conditions at $x=-L_1$ and $x=L_2$. At $x=0$, the author of
ref.~\cite{stent} imposed (i) continuity of the flux and (ii)
discontinuity in the concentration, with the concentration jump
related, via the Kedem-Katchalsky (KK) equation \cite{pharm}, to the
local flux and the membrane permeability. Explicitly, in
ref.~\cite{stent} [Eq.~(2.11) therein] the KK equation is written in
the following dimensionless form
  \begin{eqnarray}
    \frac{\partial c_2}{\partial x}=\phi\left(c_2-\frac{c_1}{\sigma}\right),
    \label{eq:KK}
    \end{eqnarray}
where $c_1$ and $c_2$ are the local concentrations on the left and
right sides of the membrane, while $\phi$ and $\sigma$ are two
transport coefficients depending on properties of the two media and
the permeability of the interface. This continuum equation can be
related to the particle-based Langevin simulations in the following
manner: Consider a small time interval small $dt$, during which a
particle residing on the left (stent) side of the membrane has a
probability $Q_1$ to cross the membrane, while a particle located on
the right (artery) side has a crossing probability $Q_2$. The crossing
probabilities are proportional to the diffusivities of the
corresponding media, and $Q_1$ also depends on the permeability of the
membrane. The flux across the membrane is given by the difference
$J=v_0[Q_1P(0-,t)-Q_2P(0+,t)]$, where $v_0$ is a proportionality
coefficient with dimensionality of velocity. Since the flux is also
given by $J=-D_1\partial_xP(0-,t)=-D_2\partial_xP(0+,t)$, we can also
write
\begin{eqnarray} 
  \frac{\partial P(0+,t)}{\partial
    x}=\frac{Q_2v_0}{D_2}\left[P(0+,t)-\frac{Q_1}{Q_2}P(0-,t)\right]
  \label{eq:langKK}
\end{eqnarray}
which has the same form as the KK equation for the boundary condition
(\ref{eq:KK}).

The results of ref.~\cite{stent} exhibit a perfect agreement with our
Langevin dynamics simulation results for $T=0$. The agreement
highlights the fact (which was not sufficiently discussed in previous
studies of the model) that the DES membrane has a nearly negligible
effect on the rate of drug flow from the stent coating to the arterial
wall. The rate of drug release from the stent can be computed by
numerical differentiation of $\Pi_1(t)$ (\ref{eq:pi1}) with respect to
time. The perfect agreement of our simulation results for $\Pi_1(t)$
with ref.~\cite{stent} indicates that the rate of drug release
decreases monotonically with time (see Fig.~6 therein).

\section{Discussion}

In this work we use underdamped Langevin dynamics simulations for
solving the diffusion equation for a simple two-layer model of a
drug-eluting stent (DES). To the best of our knowledge, this is the
first attempt to derive the solution of a DES model using this
approach. In fact, the application of underdamped Langevin dynamics
simulations is also highly uncommon in other engineering and natural
science areas dealing with the solution of complex diffusion
equations. Much more common is the use of finite element and finite
difference methods. These exist in a variety of forms, and the
selection, design, and implementation of a method that best fits a
given problem may be a complicated task. Langevin dynamics simulations
constitute an alternative approach; however, the vast majority of
studies of diffusion problems are based on simulations of the
overdamped Langevin equation which neglect the inertial term on the
r.h.s.~of Eq.(\ref{eq:langevin}). This introduces complications
stemming from the It\^{o}-Stratonovich dilemma and the associated
spurious drift. These complications can be addressed when $D(x)$ is a
smooth function which varies only slightly during an integration time
step, but not in simulations of layered systems where $D(x)$ is
discontinuous. In the latter case, one must introduce decision rules
for crossing the boundary between layers, even if those layers are
{\em not}\/ separated by a membrane. A Monte-Carlo algorithm
implementing such types of decision rules has been recently presented
\cite{slater}, but we are not aware of a similar algorithm for
overdamped Brownian dynamics. This extraordinary problem is completely
avoided in underdamped Langevin dynamics simulations which, if run
properly, produce correct thermal diffusion between sharp
interfaces. Accurate solutions for a wide range of diffusion equations
can be obtained by this approach with relative computational
simplicity and accuracy, provided that one uses a robust integration
scheme (such as the G-JF integrator) together with a suitable
convention rule for choosing the average $\alpha$ within a time-step
of the simulations (e.g., the ``inertial'' convention). The merits of
the approach are nicely exemplified in Fig.~\ref{fig:fig4b}, which
exhibits perfect agreement between the simulation results of this
paper and the predictions of ref.~\cite{stent} for the same model.

The only parameter that needs to be tuned in the simulations is the
mass of the diffusing particle whose value does not affect the PDF at
the large time scales of interest associated with the diffusion
problem. It must be selected such that the ballistic time
$\tau_m=m/\alpha$ is much smaller, but not necessarily negligible,
compared to the characteristic diffusion time. In one-dimensional
multi-layer systems, the ballistic time is set by the friction
coefficient of the most viscous layer, and the inertial convention for
the average friction needs to be applied only when the particle moves
between layers. However, the method can be readily implemented to
diffusion problems in higher dimensions and can, without any special
difficulty, be employed to study systems where the diffusion
coefficient changes continuously in space. Such problems are generally
more complicated for analytical treatment, as well as for other
computational approaches.

The model studied in this paper encompasses only two aspects of
pharmacokinetics in the DES system, namely diffusion and crossing of a
semi-permeable membrane. More recent studies suggest that advection,
caused by a pressure gradient and blood circulation, and drug binding
to receptor sites within the arterial wall may have a substantial
effect on the rate of drug transport \cite{binding}. Both mechanisms
can be dealt with within the underdamped Langevin dynamics simulation
method presented in this work. Specifically, advection results from
the action of a deterministic (``regular'') force acting on the
particle in addition to friction and thermal noise. In the presence
force, the full form of G-JF discrete-time equations must be used (see
equations (21)-(22) in~\cite{gjf1})
\begin{eqnarray}
  x^{n+1}&=&x^n+bdtv^n+\frac{bdt}{2m}\beta^{n+1}+\frac{bdt^2}{2m}f^{n},
  \label{eq:gjfx1}\\
  v^{n+1}&=&av^n+\frac{b}{m}\beta^{n+1}+\frac{dt}{2m}\left(af^{n}+f^{n+1}\right),
  \label{eq:gjfv1}
\end{eqnarray}
which include appropriate additional terms missing in equations
(\ref{eq:gjfx0}) and (\ref{eq:gjfv0}). Drug binding can be dealt with
by distributing binding sites and introducing a short-range attractive
potential between the diffusing particle and the binding sites. The
forces associated with the gradient of the binding potentials
can then be accounted for via Eqs.~(\ref{eq:gjfx1}) and
(\ref{eq:gjfv1}). In a future publication we plan to investigate model
systems including these additional effects, in order to demonstrate
the great potential of underdamped Langevin dynamics simulations for
solving non-trivial diffusion problems.

We conclude by restating that although this work deals with the
problem of drug diffusion from a DES, the larger goal of the paper is
to advocate underdamped Langeving dynamics simulations as an efficient
and simple to implement method for solving diffusion problems in
heterogeneous environments. The method is based on computation of an
ensemble of single particle stochastic trajectories with small time
steps that are in the inertial regime. This does not render the method
useless for studying problems where the time scales of interest are
well in the diffusive regime. Quite the contrary, if one is only
interested in diffusive behavior, the mass of the simulated particles
can be assigned a fictitious value (see details in section
\ref{sec:fictmass}) to allow for faster simulations with larger time
steps.

We thank MD Carlos Cafri for helpful discussions about physiological
aspects of the problem.


\begin{thebibliography}{99}
\bibitem{footnote1} The only model parameters not
  having identical values to those in {\protect \cite{stent} are the
  layers diffusion coefficients $D_1$ and $D_2$ whose values here are
  $10^3$ times smaller than in {\protect \cite{stent}}. The values
  quoted in {\protect \cite{stent}} correspond to characteristic
  diffusion times $\tau_1$ and $\tau_2$ that are unrealistically low
  (hours rather than weeks). The values of $D_1$ and $D_2$ used here
  agree with more recent estimates, see e.g.~{\protect
    \cite{review}}. This change does not affect the comparison of our
  solution for $\Pi_1(t)$ and $\Pi_2(t)$ with the data visually
  extracted from Fig.~3 in ref.{\protect ~\cite{stent}}, because the
  results throughout this paper are plotted in normalized units.}
Notice that the definition of the normalized time unit is different
here ($\tau_2=400$) than in ref.~\cite{stent} ($\tau_2=1$). The data
from \cite{stent} is plotted in Fig.~\ref{fig:fig4b} after conversion
to the time units used in this work (i.e., multiplication by a factor
of 400).
\end{thebibliography}
\end{document}